\documentclass[preprint,authoryear,12pt]{elsarticle}%
\usepackage{graphicx,longtable}
\usepackage{natbib}
\usepackage{amssymb}
\usepackage{amsthm}
\journal{New Astronomy}
\newcommand{\msun}{\mbox{\rm $M_{\odot}$~}}

\newcommand{\arcm}{\mbox{$^{\prime}$}}

\newcommand{\degree}{\mbox{$^{\circ}$}}

\newcommand{\hii}{\mbox{H{\sc ii}~}}

\def\gsim{\;\rlap{\lower 2.5pt\hbox{$\sim$}}\raise 1.5pt\hbox{$>$}\;}
\def\lsim{\;\rlap{\lower 2.5pt\hbo time lag between the formation x{$\sim$}}\raise 1.5pt\hbox{$<$}\;}
\def\la{\mathrel{\hbox{\rlap{\hbox{\lower4pt\hbox{$\sim$}}}\hbox{$<$}}}}
\def\ga{\mathrel{\hbox{\rlap{\hbox{\lower4pt\hbox{$\sim$}}}\hbox{$>$}}}}

\begin{document}
\begin{frontmatter}
\title{Pre-main-sequence population in NGC 1893 region}
\author[ari]{A. K. Pandey\corref{cor1}}
\ead{pandey@aries.res.in}
\author[ari]{M. R. Samal}
\author[ari,tw]{N. Chauhan}
\author[ari]{C. Eswaraiah}
\author[ari]{J. C. Pandey}
\author[tw]{W. P. Chen}
\author[tifr]{D. K. Ojha}
\address[ari]{Aryabhatta Research Institute of observational sciencES (ARIES), Manora Peak, Nainital, India 263 129}
\address[tw]{Institute of Astronomy, National Central University, Chung-Li 32054, Taiwan}
\address[tifr]{Tata Institute of Fundamental Research, Mumbai - 400 005, India }


\linespread{2}

\begin{abstract}
In this paper we continued our efforts to understand the star formation scenario in and around the young cluster NGC 1893. We used a sample of the young stellar sources (YSOs) identified on the basis of multiwavelength data (optical, near-infrared (NIR), mid-infrared (MIR) and X-ray) to study the nature of YSOs associated with the region. The identified YSOs show an age spread of $\sim$ 5 Myr. The YSOs located near the nebulae at the periphery of the cluster are relatively younger in comparison to those located within the cluster region. The present results are in accordance with those obtained by us in previous studies. Other main results from the present study are: 1) the fraction of disk bearing stars increases towards the periphery of the cluster; 2) there is an evidence supporting the notion that the mechanisms for disk dispersal operate less efficiently for low-mass stars; 3) the sample of Class II sources is found to be relatively older in comparison to that of Class III sources. A comparison of various properties of YSOs in the NGC 1893 region with those in the Tr 37/ IC 1396 region is also discussed. 
\end{abstract}


\end{frontmatter}

\section{Introduction}
Since it is believed that majority of the stars in the Galaxy are formed in star clusters, the star clusters and their surroundings are important tools to study the star formation process. It is believed that star formation itself is a destructive process: as massive stars form in the region, their strong stellar winds and ultra-violet (UV) radiation immediately begin to disrupt the natal environment, consequently halting further star formation in the region. Alternatively strong UV radiation from massive stars associated with ionization front (IF) can trigger next generation star formation either through `collect and collapse process', which was proposed by Elmegreen $\&$ Lada (1977) or through radiation driven implosion (RDI) of a molecular cloud condensations. Detailed model calculations of the RDI process have been carried out by several authors (e.g., Bertoldi 1989, Lefloch $\&$ Lazareff 1995, Lefloch et al. 1997, De Vries et al. 2002, Kessel-Deynet $\&$ Burkert 2003, Miao et al. 2006). An observational signature of the `collect and collapse process' is the presence of a dense layer and massive condensations adjacent to  an {\hii} region (e.g., Deharveng et al. 2003), whereas observational evidence for RDI  process is often inferred from the spatial distribution of young stars and subgroups of OB associations and their age distribution (see e.g., Matsuyanagi et al. 2006, Sharma at al. 2007, Samal et al. 2007, Pandey et al. 2008).

In the RDI process a pre-existing dense clump is exposed to the ionizing radiation from massive stars of the previous generation. The head part of the clump collapses due to  the high pressure of the ionized gas and the self-gravity, which consequently leads to the formation of next generation stars. Thus one of the signatures of the RDI process is the anisotropic density distribution in a relatively small molecular cloud surrounded by a curved ionization/shock front. Bright-rimmed clouds (BRCs) are small molecular clouds located near the edges of evolved {\hii} regions and show the above signatures. So they are considered to be good laboratories to study the physical processes involved in the RDI process. 

NGC 1893 is a very young cluster cluster (age $\sim$ 2-4 Myr) and a suitable object to study the star formation process. The cluster, located at the center of the Aur OB2 association, is associated with the {\hii} region IC 410. The cluster contains at least five O-type stars alongwith a rich population of pre-main-sequence (PMS) stars (see e.g. Sharma et al. 2007, hereafter S07). Several photometric and low resolution spectroscopic studies of the region have been carried out (for detail see S07; Prisinzano et al. 2011, hereafter P11). S07 and Maheshwar et al. (2007) have studied the star formation scenario around the NGC 1893 region. Although the studies by S07 and Maheshwar et al. (2007) reveal that stars lying away from the ionizing source are systematically younger;
 the fact manifesting a triggered star formation in the region, however, a question may be raised that the observed age sequence may be biased as those studies had used mainly classical T-Tauri stars (CTTSs) which either show a significant amount of H$\alpha$ emission and/ or NIR excess. On the basis of NIR $(J-H)/ (H-K)$ CC diagram the identification of weakline T-Tauri stars  (WTTSs) and CTTSs with small NIR excess is difficult as the NIR-CC diagram is significantly contaminated by the field star population (see e.g. Figure 16 of S07).

Recently, MIR $Spitzer$/IRAC and X-ray data from {\it Chandra} telescope have become available (cf. Caramazza et al. 2008, P11). Therefore, we considered worthwhile to re-investgate the star formation scenario in/ around of NGC 1893 region as well as to study the evolution of disk of TTSs in the NGC 1893 region using the newly available data.

\section {Data}

On the basis of optical, NIR, MIR and X-ray data P11 have identified 1034 and 442 Class II and Class III YSOs, respectively. We have used the optical and NIR data for the above mentioned YSOs given by P11. Whenever the NIR data for these sources are not available in the catalogue by P11, the same have been taken from the Two Micron All Sky Survey (2MASS) Point Source Catalog (PSC) (Cutri et al. 2003). Sources having uncertainty $\leq$ 0.1 mag (S/N $\geq$ 10) in all the three bands were selected to ensure high quality data. The $JHK_s$ data were transformed from the 2MASS system to the California Institute of Technology (CIT) system using the relations given in the 2MASS website.  

\section {Distance to the cluster}

\begin{figure*}
\resizebox{12cm}{12cm}{\includegraphics{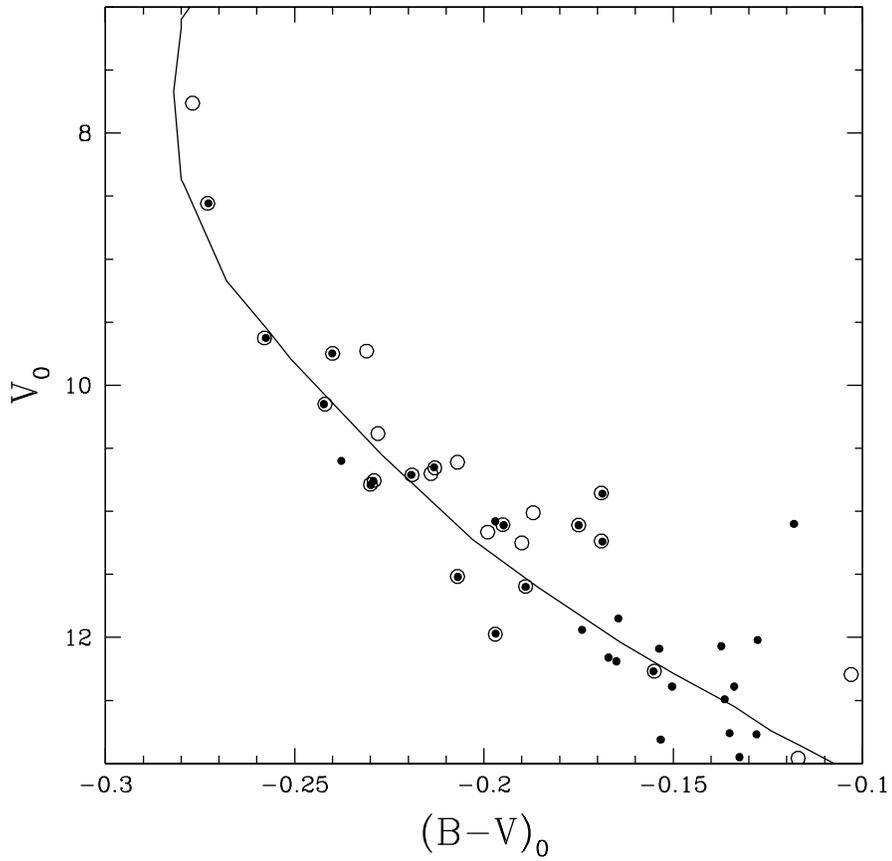}}
\caption{ $V_0/ (B-V)_0$  CMD for stars lying within  $r \le 3^{\prime}$ of NGC 1893 and having spectral type earlier than A0 (filled circles). The open circles are  the probable cluster  members (lying within $ r \le 10^{\prime}$)  identified by Eswaraiah et al. (2011) on the basis of polarimetry and photometry.  The isochrone of 4 Myr age (continuous line) by Girardi et al. (2002) corrected for the cluster distance is also shown. }
\label{fig1}
\end{figure*}

S07 estimated a variable reddening ($E(B-V) = 0.4 - 0.6$ mag) towards the direction of the cluster. Assuming a foreground reddening of $E(B-V) = 0.4$ mag, they estimated the distance to the cluster as 3.25 kpc using the $V/ (V-I)$ colour-magnitude diagram (CMD). P11 using a mean reddening $E(B-V) = 0.6$ and $V/ (U-B)$ CMD obtained a slightly higher value (3.6 kpc) for the distance. In our opinion the use of mean value of reddening is not appropriate to estimate the distance as it will yield a higher value to the distance. 

Here we have re-estimated the distance to the cluster using the dereddened $V$ mag and $(B-V)$ colour. To avoid the contamination due to field stars we used the stars lying within 3 arcmin radius of the cluster. Reddening of individual stars having spectral types earlier than A0 has been computed by using the reddening free index $Q$ (Johnson \& Morgan 1953). Assuming a normal reddening law we can construct a reddening-free parameter index $Q = (U-B) - 0.72\times (B-V)$.
The value of $Q$ for stars earlier than A0 will be $< 0$. For main-sequence (MS) stars, the intrinsic $(B-V)_0$ colour and colour-excess can be obtained from the relation $(B-V)_0 = 0.332\times Q$  (Johnson 1966; Hillenbrand et al. 1993) and $E(B-V) = (B-V) - (B-V)_0$, respectively. Fig. 1 shows dereddened $V_0/(B-V)_0$  CMD for stars lying within 3 arcmin of the cluster center alongwith the probable members in the NGC 1893 cluster region ($r \le 10^{\prime}$, open circles) identified by Eswaraiah et al. (2011) on the basis of polarimetric and photometric observations.  
 Using the theoretical isochrone of 4 Myr (Z = 0.02; log age = 6.6 yr) by Girardi et al. (2002) we estimated a distance of $3.3 \pm 0.4$ kpc which is in agreement with the distance (3.25 kpc) obtained by S07. Hence in the present study we use the distance to the cluster as $3.3 \pm 0.4$ kpc.

\section {Pre-main-sequence members in the region}
\begin{figure*}
\resizebox{12cm}{12cm}{\includegraphics{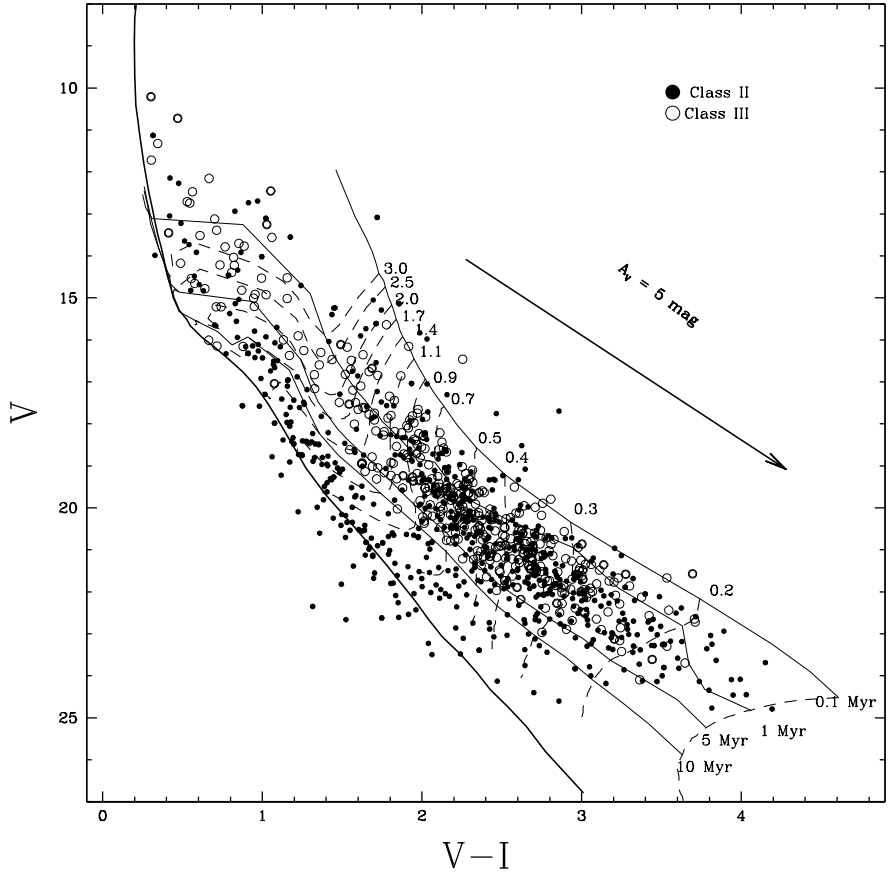}}
\caption{$V,(V-I)$ CMD for the YSOs identified by P11 in the NGC 1893 region. 
The thick curve is the isochrone for 4 Myr from Girardi et al. (2002). Thin 
curves are the PMS isochrones for 0.1, 1, 5 and 10 Myr by Siess et al. (2000). 
The dashed curves represent evolutionary tracks for different masses by Siess 
et al. (2000). All the isochrones and tracks are corrected for the distance 
and reddening. The filled and open circles represnt Class II and Class III 
sources respectively. The arrow shows the reddening vector.}
\label{fig2}
\end{figure*} 

Fig. 2 shows $V/(V-I)$ CMD for the optical counterparts of the YSOs  (646 Class II and 425 Class III sources) identified by P11. In Fig. 2 we have also plotted post main sequence isochrones of 4 Myr with solar metallicity by Girardi et al. (2002) as well as pre-main-sequence isochrones by Siess et al. (2000). All the isochroes are adjusted for the cluster distance of 3.3 kpc and reddening $E(B-V)$ = 0.40 mag. Almost all the Class III sources have ages $\lesssim$ 5 Myr, whereas a significant number of Class II sources, as also pointed out by P11, are located below the 5 Myr PMS isochrone. P11 discussed that these could be candidate members with anomalous optical colours and/ or magnitude due to the disk related phenomena. In the case of W5 E {\hii} region ($l= 137\degree.2$, $b= 0\degree.92$) Chauhan et al. (2011) have found a significant amount of contamination in the sample of YSOs identified by Koenig et al. (2008) on the basis of {\it Spitzer} data. Chauhan et al. (2011) used $(J-H)/ (H-K)$  colour - colour diagram to avoid the contamination due to field star population in the YSO sample selected on the basis of {\it Spitzer} data.  Since NGC 1893 is located in the galactic plane, the cluster region is expected to be significantly contaminated by foreground as well as background stars (see S07). To understand the star formation scenario in the region, it is necessary to consider only those stars which are associated with NGC 1893 region. To avoid the contamination in the sample of YSOs given by P11 we adopted the following 
approach.

\subsection {Identification of pre-main sequence stars}
Figs 3a (left panel) and 3b (right panel) display NIR colour - colour (NIR- CC) diagrams for the Class II and Class III sources identified by P11. In Fig. 3 the thin and thick dashed curves represent the unreddened main-sequence and giant branches (Bessell $\&$ Brett 1988), respectively. The dotted line indicates the locus of intrinsic CTTSs (Meyer et al. 1997). The curves are in the CIT system. The parallel dashed lines are the reddening vectors drawn from the tip of the giant branch (``upper reddening line''), from the base of the main-sequence (MS) branch (``middle reddening line'') and from the base of the tip of the intrinsic CTTSs line (``lower reddening line''). The extinction ratios $A_J/A_V = 0.265, A_H/A_V = 0.155$ and $A_K/A_V=0.090$ have been adopted from Cohen et al. (1981). The NIR-CC diagram is classified into three regions (cf. Ojha et al. 2004a,2004b). `F' region is located between the upper and middle reddening lines and the sources lying in this region could be either field stars (main-sequence stars, giants) or Class III and Class II sources with small NIR excesses. The `T' region is located between the middle and lower reddening lines.  The `T' region sources are considered to be mostly CTTSs (Class II objects). There may be an overlap in NIR colours of Herbig Ae/Be stars and CTTSs in the `T' region (Hillenbrand et al. 1992). The `P' region is redward of the `T' region and the sources lying in this region are  likely Class I objects (protostar-like objects; Ojha et al. 2004b). Therefore  objects falling in the `T' and `P'  regions of NIR-CC diagram could be probable Class II members associated with the cluster region. It is worthwhile, however, to mention that Robitaille et al. (2006) have recently shown that there is a significant overlap between protostars and CTTSs in the NIR-CC space.

\begin{figure*}
\resizebox{6cm}{6cm}{\includegraphics{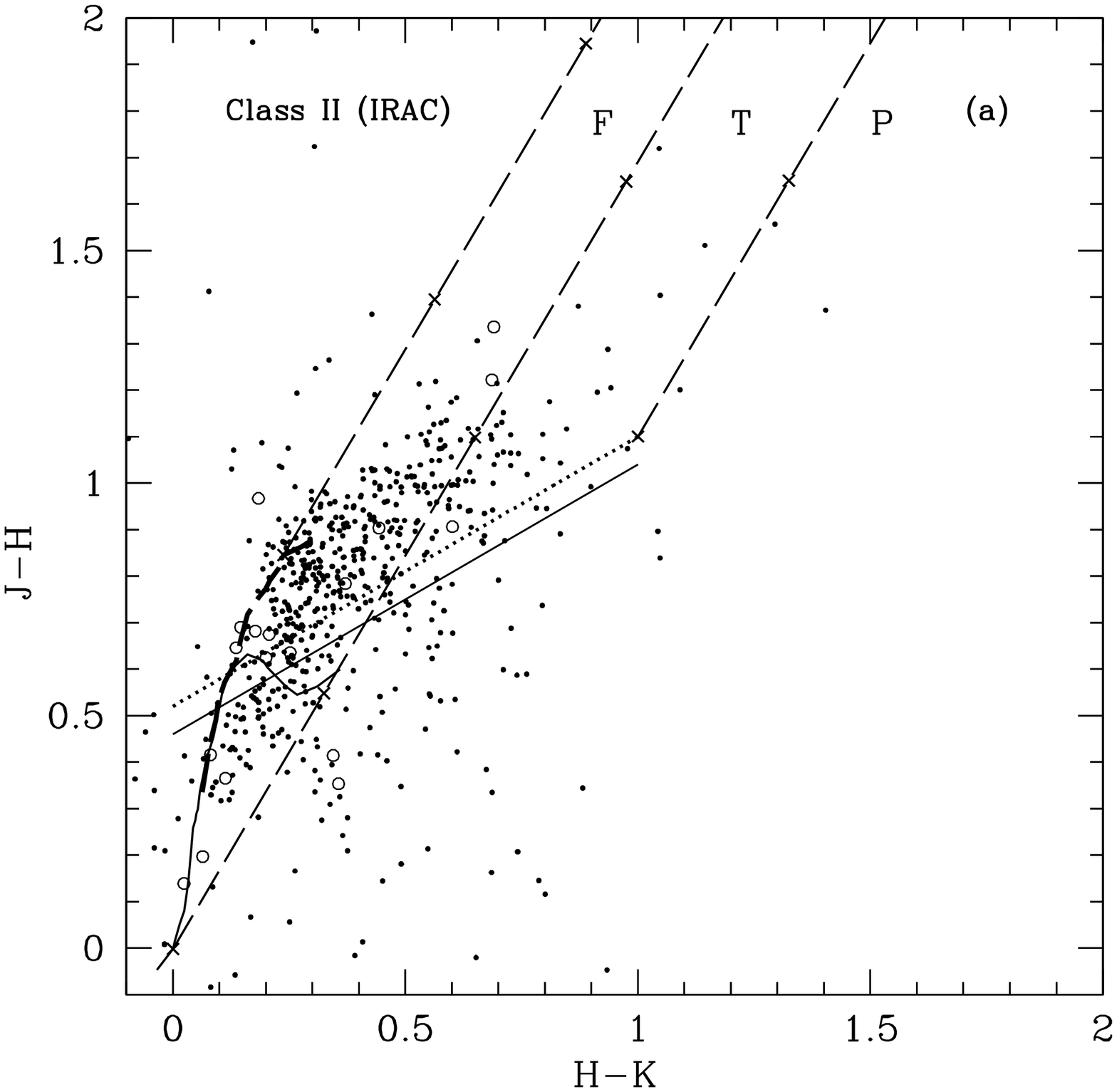}}
\resizebox{6cm}{6cm}{\includegraphics{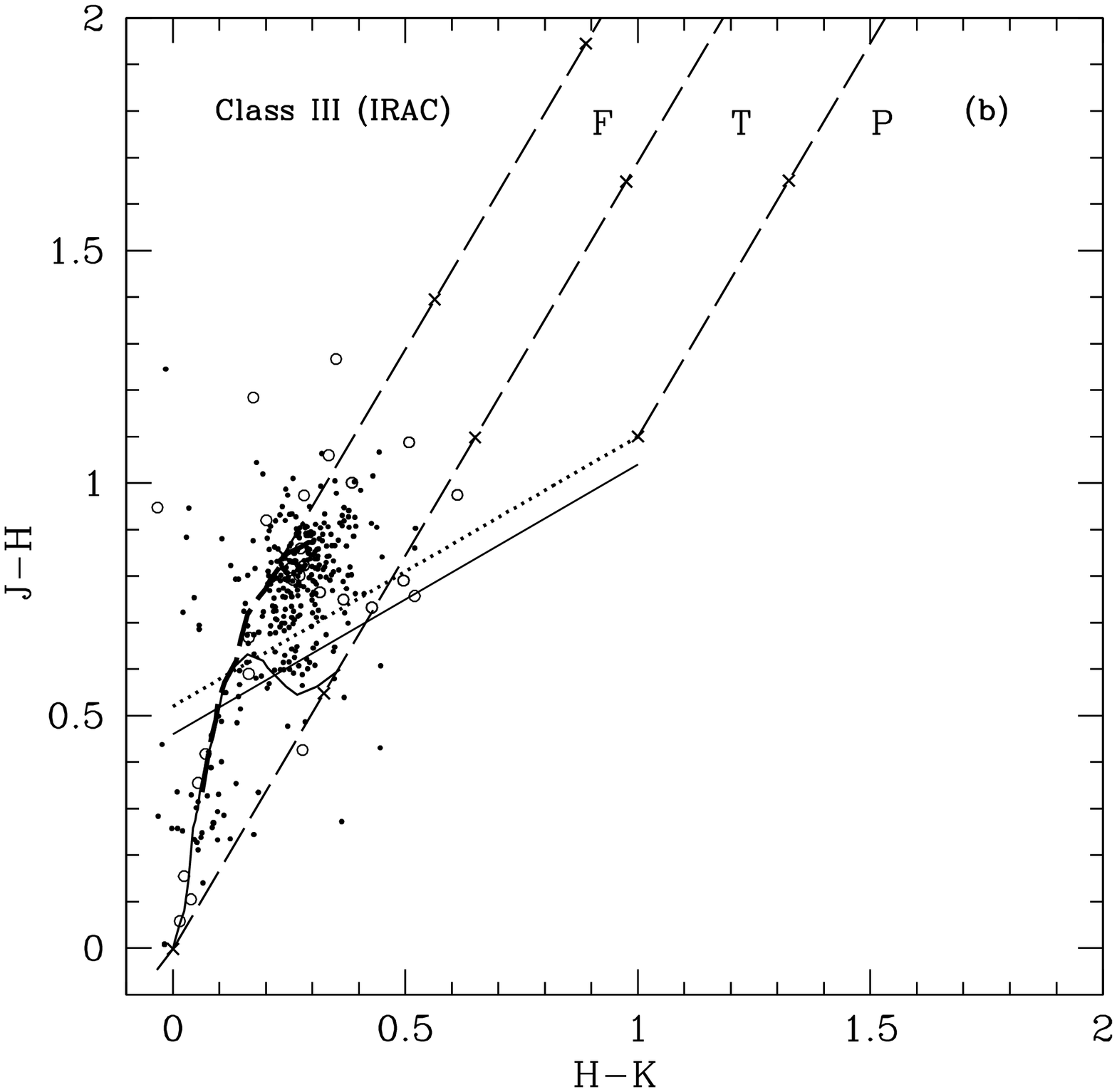}}
\caption{(a) NIR CC diagram for all the sources classified as Class II sources 
in the catalogue by P11. The $JHK$ data for open circles have been taken from 
2MASS catalogue. The thin and thick dashed curves represent the unreddened 
main sequence and giant branches (Bessell $\&$ Brett 1988), respectively. 
The dotted line indicates the locus of intrinsic CTTSs (Meyer et al. 1997).  
The curves are also in the CIT system. Considering the errors in the data, we 
consider only those sources as PMS sources associated with the region which 
lie above the continuous line. The parallel dashed lines are the reddening 
vectors drawn from the tip of the giant branch (``upper reddening line''), from the base of the main-sequence branch (``middle reddening line'') and from the 
base of the tip of the intrinsic CTTS line (``lower reddening line''). (b) Same as Fig. 3a but for Class III sources.  }
\label{fig3}
\end{figure*}

Fig. 3a reveals that a large number of sources classified as Class II lie below the intrinsic locus of CTTSs. The sources having $(H-K) \lesssim$ 0.3 mag and lying below the extension of the intrinsic CTTSs locus could be reddened ($A_V \sim 2 -4$ mag) background MS stars. Fig. 3b also shows that a few Class III sources having $(J-H) \lesssim$ 0.5 mag are located around the unreddened MS locus. The location of these sources in $V/ (V-I)$ CMD also corresponds to the upper MS locus with magnitude brighter than $V\sim $15 mag (cf. Fig. 2). Damh \& Simon (2005) 
have identified CTTSs and WTTSs in NGC 2264 region on the basis of the 
H$\alpha$ sources. A comparison of NIR-CC diagrams for the Class II and Class III sources of NGC 1893 (Fig. 3) with the NIR-CC diagrams of CTTSs and WTTSs in the NGC 2264 region (cf. figure 16 by Dahm \& Simon 2005) manifests that the sample of YSOs identified by P11 is stongly contaminated by field populations. In Fig. 3 we plot a continuous line parallel to the intrinsic CTTS locus. Considering the errors in the data, we consider only those sources as PMS sources associated with the region which lie above the continuous line.  The above criteria yield 367 and 246 Class II and Class III sources, respectively with optical counterparts.

\section {AGE AND MASS ESTIMATION}

\begin{figure}
\centering
\resizebox{12cm}{12cm}{\includegraphics{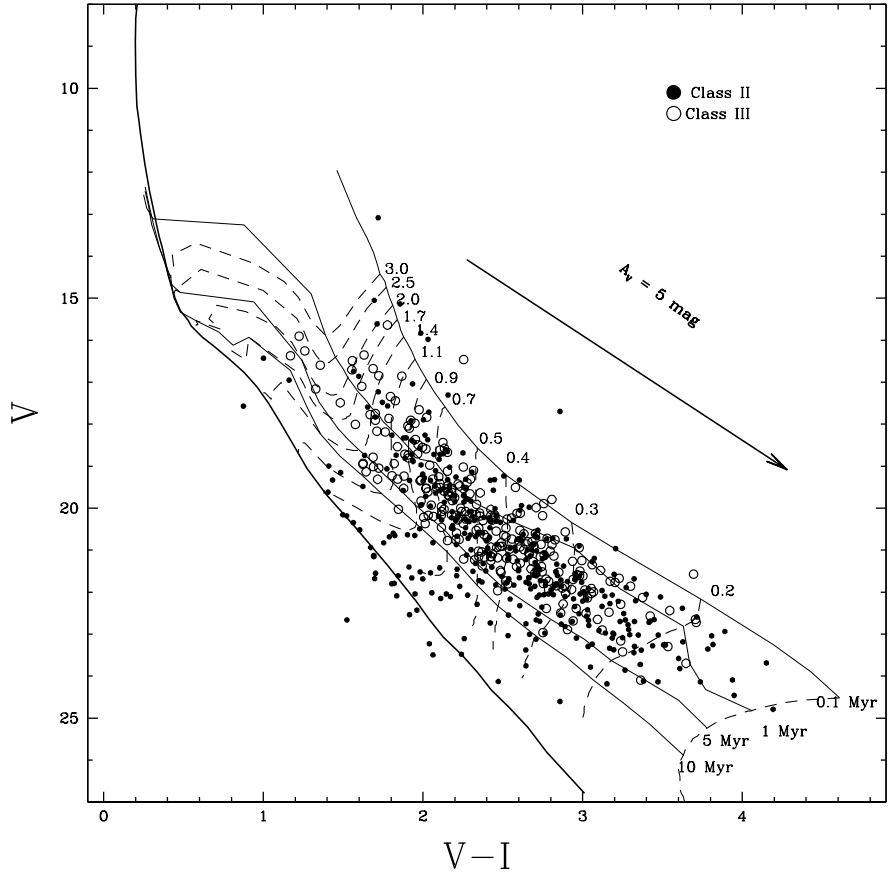}}
\caption{ $V$ vs $V-I$  CMD of the selected YSOs. The isochrone for 4 Myr age (thick solid line) by Girardi et al. (2002) and PMS isochrones of 0.1, 1, 5 and 10 Myr (thin lines)  along with evolutionary tracks (dashed lines) of different mass stars by Siess et al. (2000) are also shown. All the isochrones are corrected for a distance of 3.3 kpc and reddening $E(B-V)=0.4$ mag. The symbols are same as in Fig. 2. }
\label{fig4}
\end{figure}

Fig. 4 shows $V/ (V-I)$ CMD for all the selected YSOs in the region as mentioned in Sec 4. In Fig. 4 we have also plotted the isochrone for 4 Myr by Girardi et al. (2002), which practically represents the zero-age-main-sequence (ZAMS), along with  the PMS isochrones for 0.1, 1, 5, 10 Myr with solar metallicity by Siess et al. (2000). The distance and  reddening $E(B-V)_{min}$ have been taken as 3.3 kpc and 0.4 mag (cf. Sec 3 and S07). Fig. 4 manifests that the contamination due to non-members is greatly reduced by applying the selection criterion as mentioned in Sec. 4. A similar case has been observed in W5 E region by Chauhan et al. (2011). The CMD (Fig. 4) shows a fairly well similarity with the statistically cleaned CMD as shown by S07 and also reveals an age spread in the cluster region.
Although the contamination is significantly reduced, still we can notice a few sources classified as Class II sources by P11 below the 5 Myr PMS isochrone. The CMD for TTSs identified on the basis of H$\alpha$ emission stars in the NGC 2264 region (Dahm \& Simon 2005) and in the IC 1396 region (Barentsen et al. 2011) shows a significant number of sources below 10 Myr isochrone. Barentsen et al. (2011) have discussed these sources and concluded that these could be background objects. To avoid the contamination due to background stars, we have considered only those sources as the PMS objects associated with NGC 1893 which are located above the 5 Myr isochrone.  This selection yields 298 and 240 Class II and Class III sources, respectively.

 The age and mass of each YSO, given in Table 1\footnote{The complete Table is available only in electronic form.}, are estimated by using the $V, (V-I)$ CMD, as discussed by Pandey at al. (2008) and Chauhan et al. (2009). Here we would like to point out that the estimation of the ages and masses of the PMS stars by comparing their locations in the CMDs with the theoretical isochrones is prone to random as well as systematic errors (see e.g. Hillenbrand 2005, Hillenbrand 2008, Chauhan et al. 2009, 2011). The effect of random errors due to photometric errors and reddening estimation in determination of ages and masses was estimated by propagating the random errors to their observed estimation by assuming normal error distribution and using the Monte-Carlo simulations (see e.g, Chauhan et al. 2009). The estimated ages and their errors are also given in Table 1. 
The systematic errors could be due to the use of different PMS evolutionary models and the error in distance estimation etc. Since we are using evolutionary models by Siess et al. (2000) to estimate the age of all the YSOs in the region, we presume that the age estimation is affected only by the random errors.
The presence of binaries may be the another source of error in the age determination. Binarity will brighten the star, consequently the CMD will yield a lower age estimate. In the case of an equal mass binary we expect an error of $\sim$ 50 - 60\% in the PMS age estimation. However, it is difficult to estimate the influence of binaries/variables on mean age estimation as the fraction of binaries/variables is not known. In the  study of TTSs in the {\hii} region IC 1396, Barentson et al. (2011) presumed that the number of binaries in their sample of TTSs could be very low as close binary lose their disc significantly faster than single stars (cf. Bouwman et al. 2006).
Estimated ages and masses of the YSOs ranges from $\sim $ 0.1 to 5 Myr and $\sim $ 0.1 - 3.0 $M_\odot$ respectively, which are comparable with the lifetime and masses of TTSs.

\section {STAR FORMATION AND DISK EVOLUTION}
Massive O-type stars may have strong influence and significantly affect the entire star forming regions.  Energetic stellar winds from massive stars, on one hand, could evaporate nearby cloud and consequently terminate  star formation. Alternatively stellar winds and shock waves  may squeeze molecular cloud clumps and induce subsequent star formation. The morphological details of the environment and distribution of the YSOs in the NGC 1893 region can be used to probe the star formation scenario in the region. Here we felt worthwhile to compare the star formation scenario in the NGC 1893 region with that in the IC 1396 star forming region located at a distance of 870 pc and has rather similar morphology like NGC 1893. 

\begin{figure*}
\resizebox{6cm}{6cm}{\includegraphics{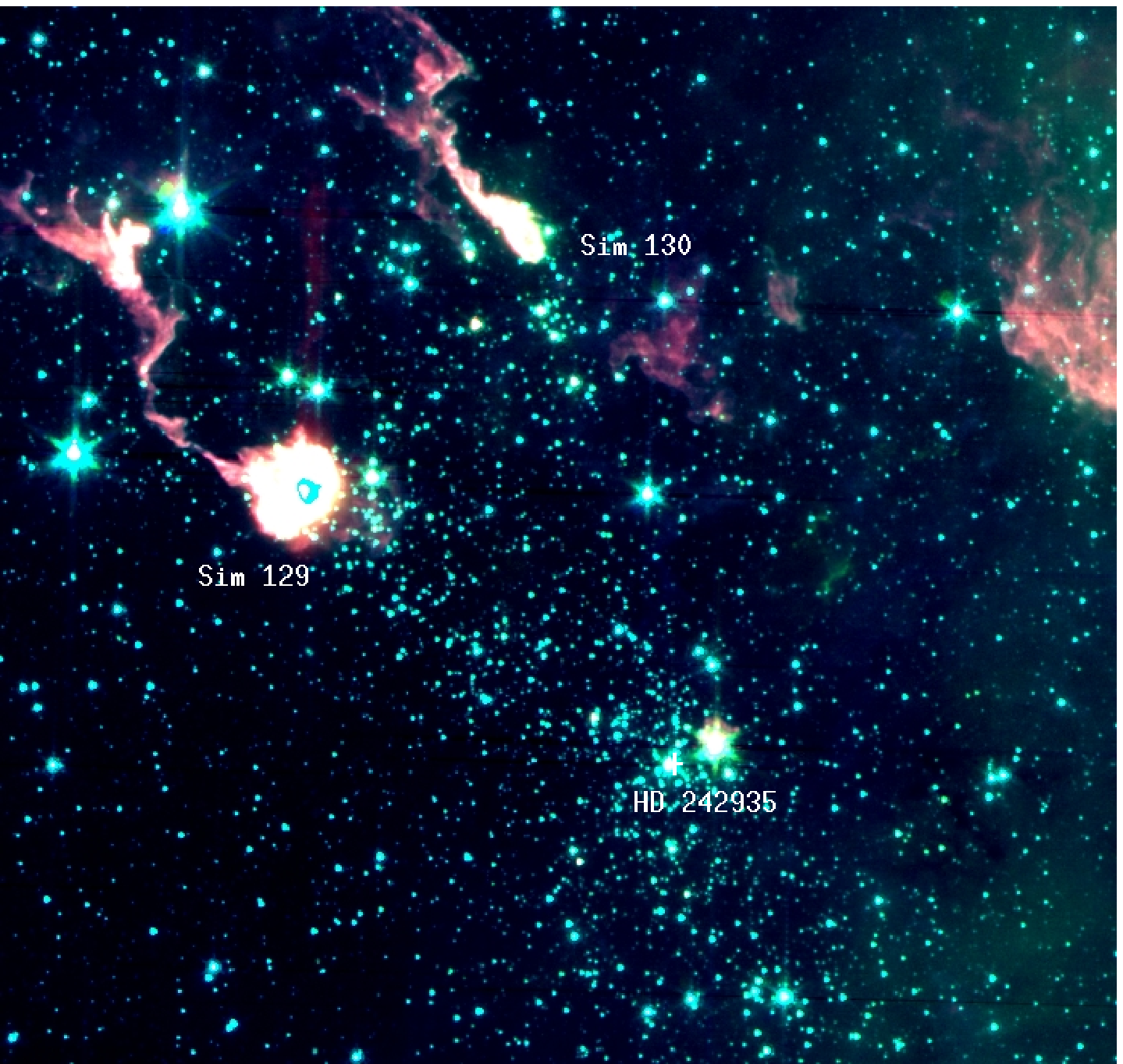}}
\resizebox{6cm}{6cm}{\includegraphics{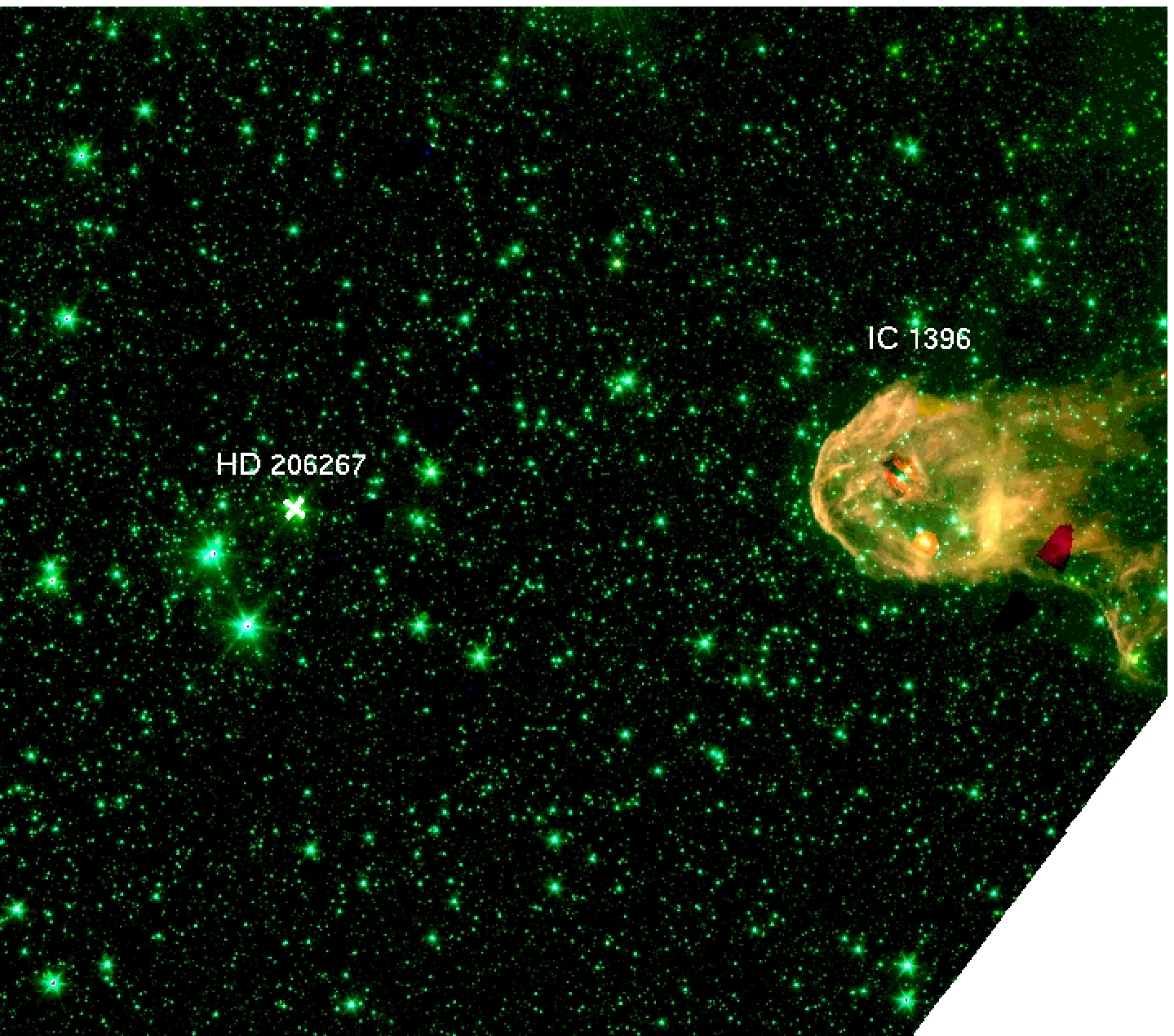}}
\caption{ False color composite image using the {\it Spitzer} observations (blue:3.6 $\mu$m, green:4.5 $\mu$m, red:8.0 $\mu$m) of NGC 1893 (left panel) and IC 1396 (right panel) regions. North is up, and East is to the left. }
\label{fig5}
\end{figure*}

\subsection {Morphology}

Figure 5 shows  colour composite images (blue: 3.6 $\mu$m, green: 4.5 $\mu$m, red: 8.0 $\mu$m) from {\it Spitzer} observations for NGC 1893 and IC 1396 regions. Both the regions, NGC 1893 and IC 1396, have globules  at a radial distance of $\sim$ 5.5 pc and  $\sim$ 4.5 pc respectively from the cluster centre. The ages of these clusters are $\lesssim$ 4 Myr and both clusters contain  massive O-type stars  at the center those regulates the star formation activity through stellar radiation and/ or wind (S07, Sicilia-Aguilar et al. 2005). Although NGC 1893 contains five O-type stars and 12 B1-type stars at various locations in the complex, the region is mainly powered by an O7.5 (HD 242935) star at the cluster center. 
The IC 1396 region contains cluster Tr 37 with two O-type and 12 stars earlier than B1-type. The region is mainly powered by an O6 star (HD 206267) which is a member of the cluster Tr 37 and has several globules as well as BRCs at its periphery. In the case of IC 1396, it is believed that an expanding IF might have interacted with the molecular cloud and presently seen as globules and BRCs. NGC 1893 has a high density region on the south-west side and low density matter towards the north-east direction  which allows the matter to flow thus producing a typical "champagne flow" morphology (cf. Maheswar et al. 2007). The two pennant nebulae Sim 129 and Sim 130 are situated in the champagne flow direction. 

Chen et al. (2004) and Sharma et al. (2006) have discussed that the initial stellar distribution in star clusters may be governed by the structure of parental molecular cloud as well as how star formation proceeds within the cloud. Later evolution of the cluster may be governed by internal gravitational interaction among member stars and external tidal forces due to the
Galactic disc or giant molecular clouds. S07 have studied the morphology of the cluster NGC 1893 on the basis of isodensity contours generated from optical ($V\le 18$) as well as from NIR 2MASS data. The isodensity contours indicate that the cluster NGC 1893 has elongated morphology. In Fig. 6 we have plotted the isodensity contours for the YSO sample identified in Section 4.1. The contours are plotted above  $1\sigma$ level. The distribution of YSOs indicates that the cluster NGC 1893 has elongated morphology which agrees well with the morphology obtained by S07. Since the cluster is not dynamically relaxed (S07), the present morphology of the cluster NGC 1893 may be governed by the star formation process.

\begin{figure*}
\resizebox{12cm}{12cm}{\includegraphics{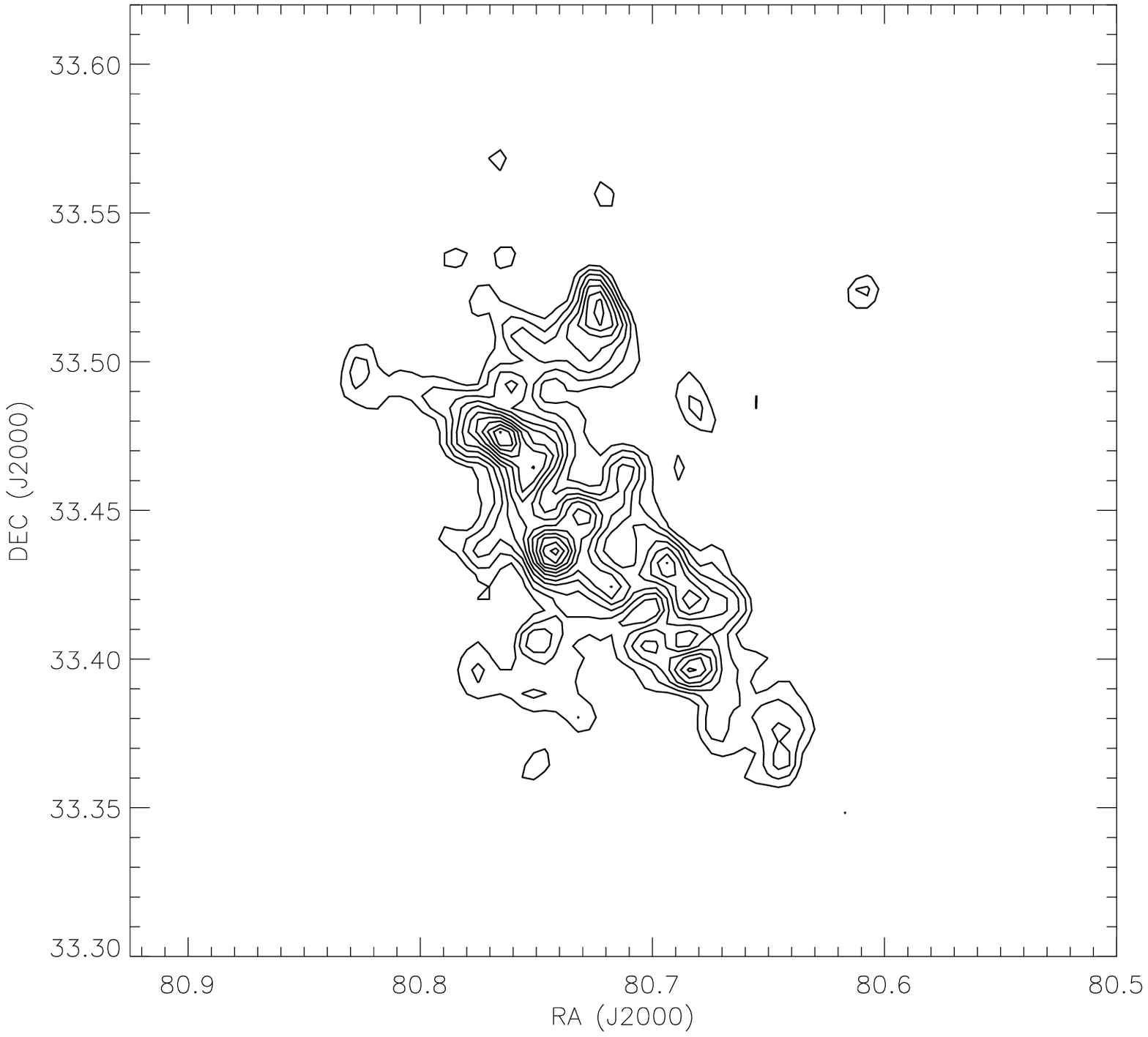}}
\caption{ Isodensity contours for the YSOs identified in the present study. The contours are plotted above $1\sigma$ level. The contours have step size of 2 stars/ arc-min$^{2}$ with the lowest contour representing 4 stars/ arc-min$^{2}$. The maximum stellar density is 40 stars/ arc-min$^{2}$ . North is up and East is to the left. }
\label{fig6}
\end{figure*}

\subsection {Spatial distribution of YSOs}

\begin{figure*}
\resizebox{6.2cm}{6cm}{\includegraphics{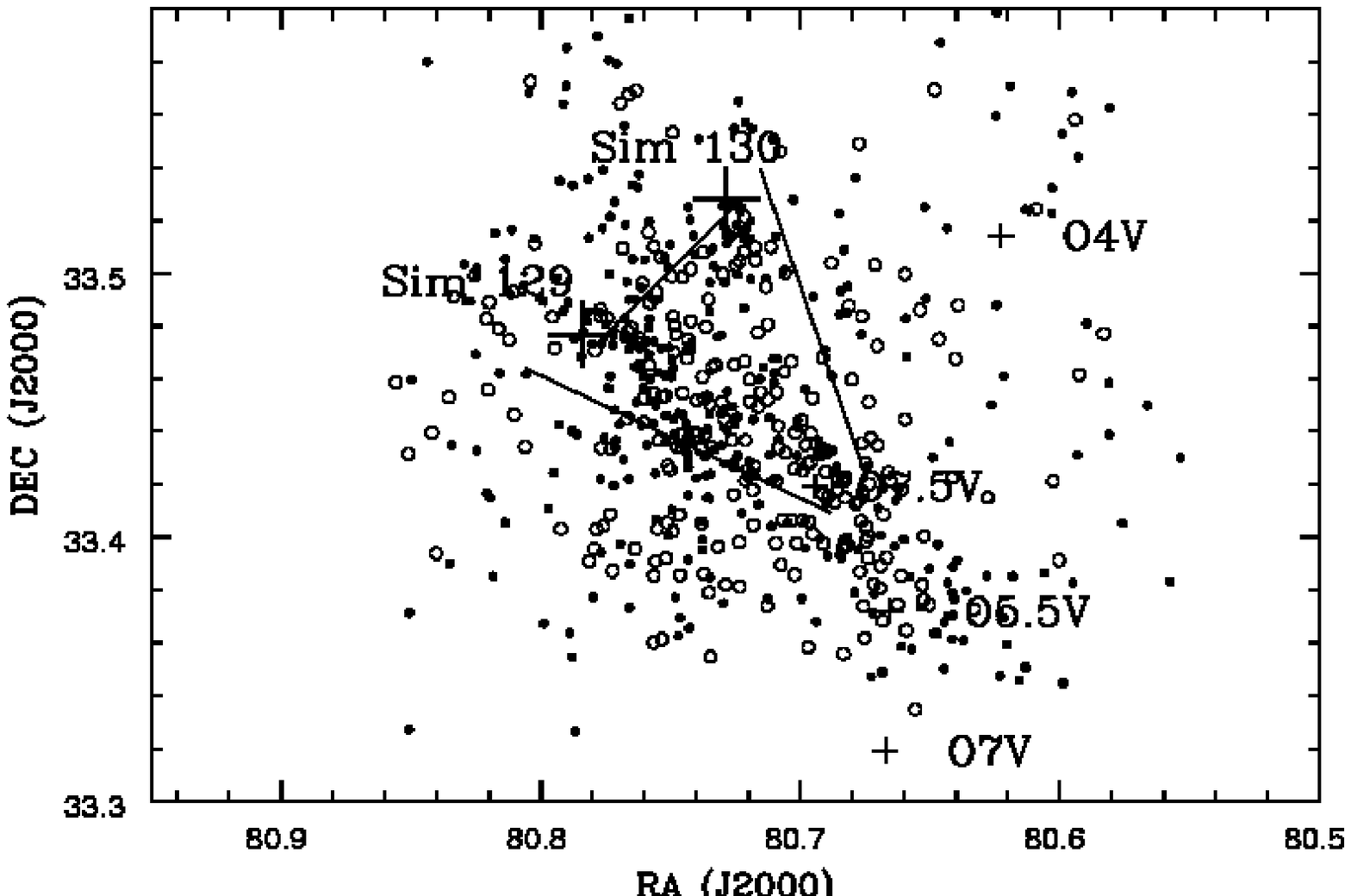}}
\resizebox{6.3cm}{6cm}{\includegraphics{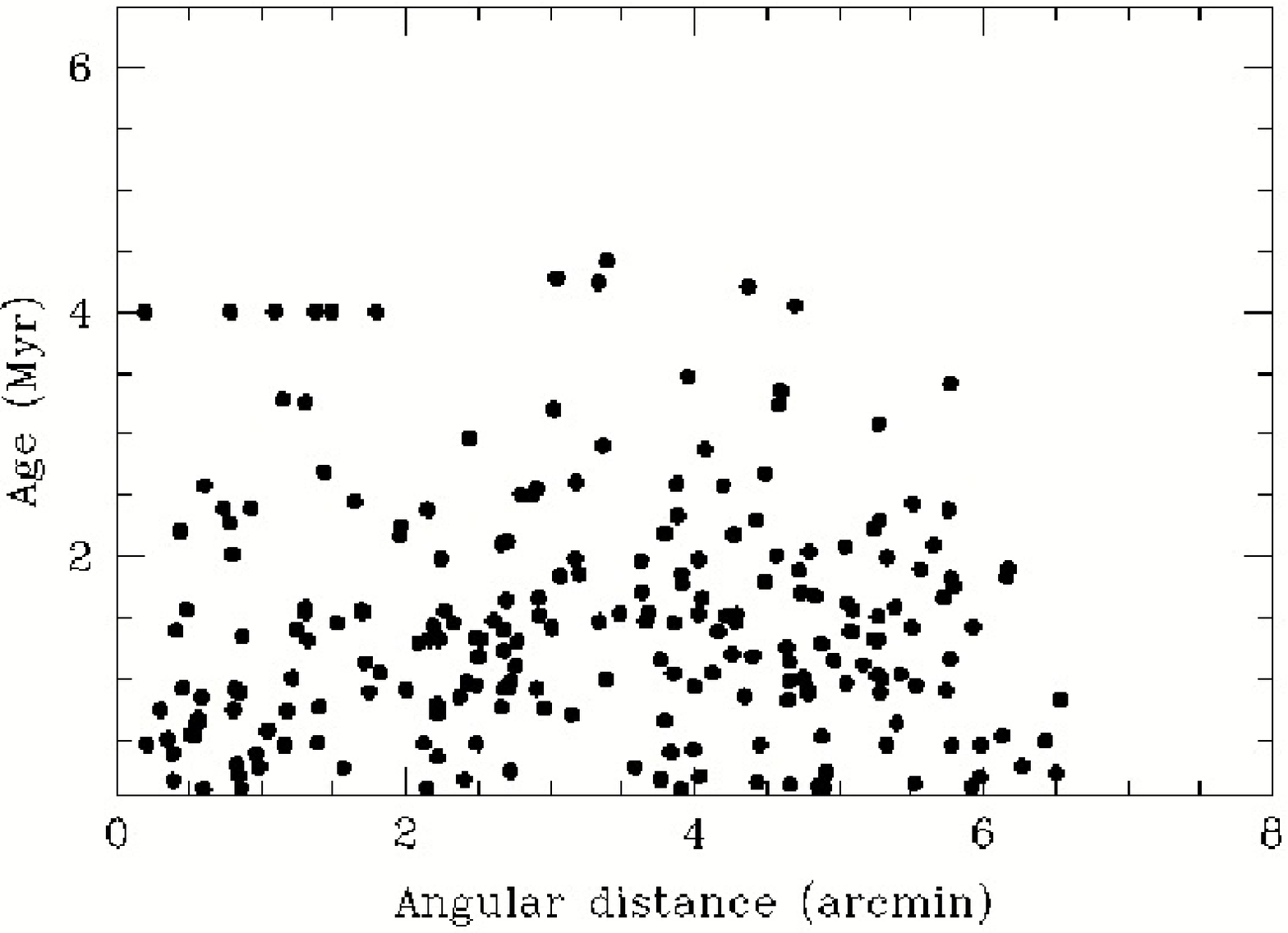}}
\caption{ Left Panel: The distribution of selected YSOs (Class II: filled  circle, Class III: open circles) in NGC 1893 region. The positions of the emission nebulae Sim 129 and Sim 130 are marked with large plus sign. The distribution of O-type stars with their spectral types is marked with small plus signs.
The sources bounded by the straight lines have  been used to study the age distribution of the YSOs (see right panel).  Right panel: The  age distribution of YSOs lying within the marked region (see left panel) as a function of radial distance from  the ionising source HD 242935 of NGC 1893 region.}
\label{fig7}
\end{figure*}

On the basis of distribution of H$\alpha$ emission and NIR excess stars S07 and Maheswar et al. (2007) have argued that star formation in the region have taken place through the RDI process. Fig. 7 (left panel) shows spatial distribution of all the selected YSOs in the NGC 1893 region as described in Sec 4.
Fig. 7 (left panel) reveals a rather aligned distribution of YSOs from the ionization source to the direction of nebulae as reported by Maheswar et al. (2007). Fig. 7 (right panel) displays the age distribution of YSOs lying within the region bounded by continuous lines in the left panel as a function of radial distance from the ionization source HD 242935, which manifests a scatter in the ages ($\sim$ 3-4 Myr) of YSOs lying in the  cluster region (r $\lesssim$ 5 arcmin), whereas  YSOs located near the nebulae at the periphery of the cluster have rather smaller ages in comparison to the YSOs lying in the cluster region. This further supports the notion of triggered star formation in the region due to RDI process as suggested by Maheswar et al. (2007) and S07. In the case of the Tr 37/ IC 1396 globule region Sicilia-Aguilar et al. (2004) 
have shown that the H$\alpha$ emission stars (i.e., TTSs) are found to be aligned towards the direction of IC 1396 globule from the ionizing source, HD 206267 (O6 star). They also reported that most of the younger ($\sim$1 Myr) members appear to lie near or within the IC 1396 globule and concluded that it can be indicative of the triggered star formation. The age distribution of YSOs located between Tr 37 and IC 1396 globule as a function of radial distance from the star HD 206267 is shown in Fig. 8, which reveals a distribution rather similar to that shown in Fig. 7 (right panel). The ages of the YSOs have been taken from Sicilia-Aguilar et al. (2004). The ages of the YSOs located at $r \gtrsim$ 12 arcmin have relatively 
lower ages. A similar trend has recently been found for YSOs associated with the NGC 281 region (Sharma et al. 2012). In the case of a few BRCs Chauhan et al. (2009) have also reported that the  H$\alpha$ emission and NIR excess stars show an aligned distribution from the ionizing source to the direction of bright rims.

\begin{figure}
\centering
\resizebox{12cm}{12cm}{\includegraphics{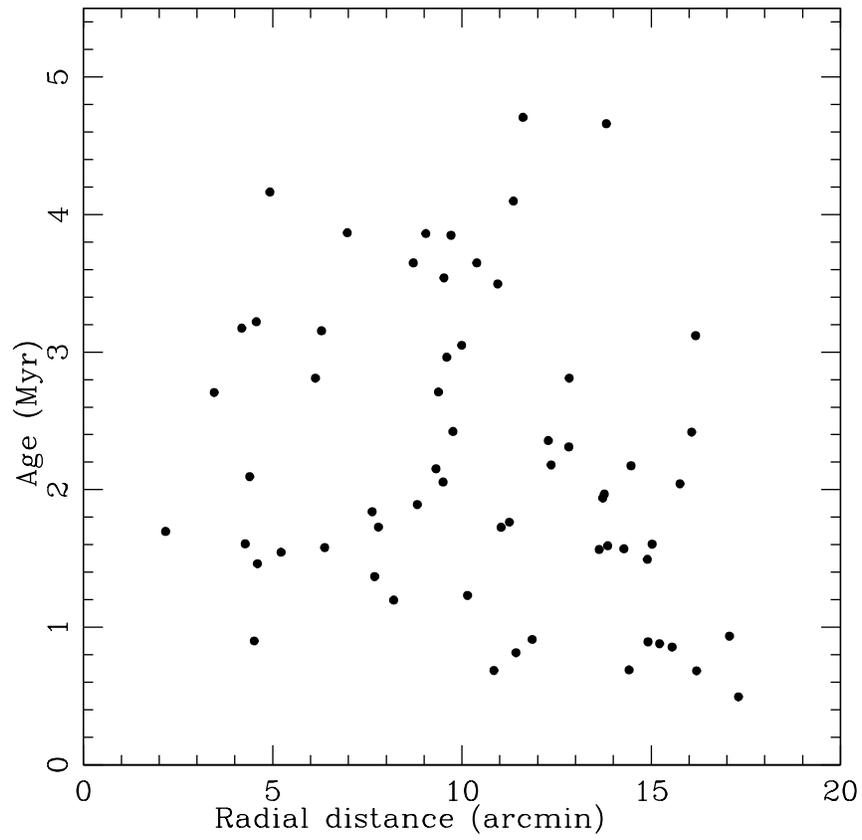}}
\caption{Age distribution of YSOs towards the  direction of IC1396 globule as a function of radial distance from the O6.5 ionizing star HD 206267 of Tr37.}
\label{fig8}
\end{figure}

Figure 9 compares the radial distribution of YSOs in the NGC 1893/ Sim 129 \& 130  and Tr 37/ IC 1396 regions. The upper and lower insets in Fig. 9 display the fraction of CTTSs (assuming that all Class II sources and Class III sources are CTTSs and WTTSs respectively), $f_{CTTS} = (N_{CTTS}/ N_{CTTS}+N_{WTTS})$ towards the direction of the nebulae  and for the whole cluster  region respectively, as a function of projected radial distance from the ionization source. The fraction $f_{CTTS}$ shows an increasing trend as we move away from the ionization source. In the case of NGC 2244, Balog et al. (2007) have found that the fraction of stars with disk does not correlate with the distance, however they found a deficit of disks within the 0.5 pc radius from the O stars. Johnstone et al.(2004) have reported that the far-UV radiation from nearby massive star(s) may cause photoevaporation of YSO disks resulting in short ($\sim$ 10$^{6}$ yr) disk lifetimes. In the case of IC 1396 Barentsen et al. (2011) have concluded that the spatial gradient of IR excess is unlikely to be explained by photoevaporation alone. Roccatagliata et al. (2011) in the case of IC 1795 association do not find any systematic variation in the spatial distribution of disks within 0.6 pc of O-type star. The theoretical calculations predict that external UV radiation due to high-mass star can photoevaporate outer disks only within 0.3-0.7 pc (see e.g., Johnstone et al. 1998, Adams et al. 2004, Clarke 2007).  

\begin{figure*}
\resizebox{6cm}{6cm}{\includegraphics{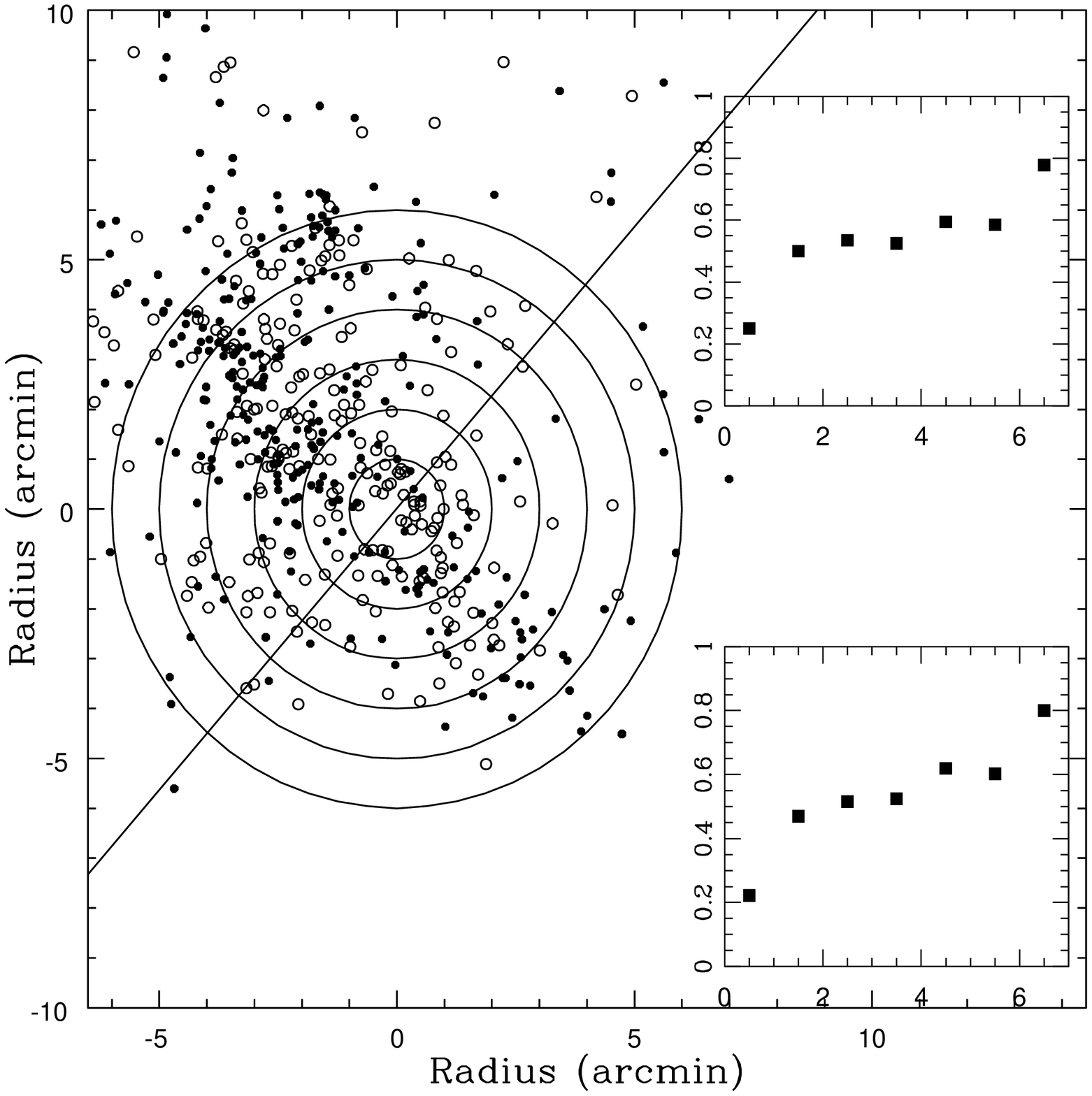}}
\resizebox{6cm}{6cm}{\includegraphics{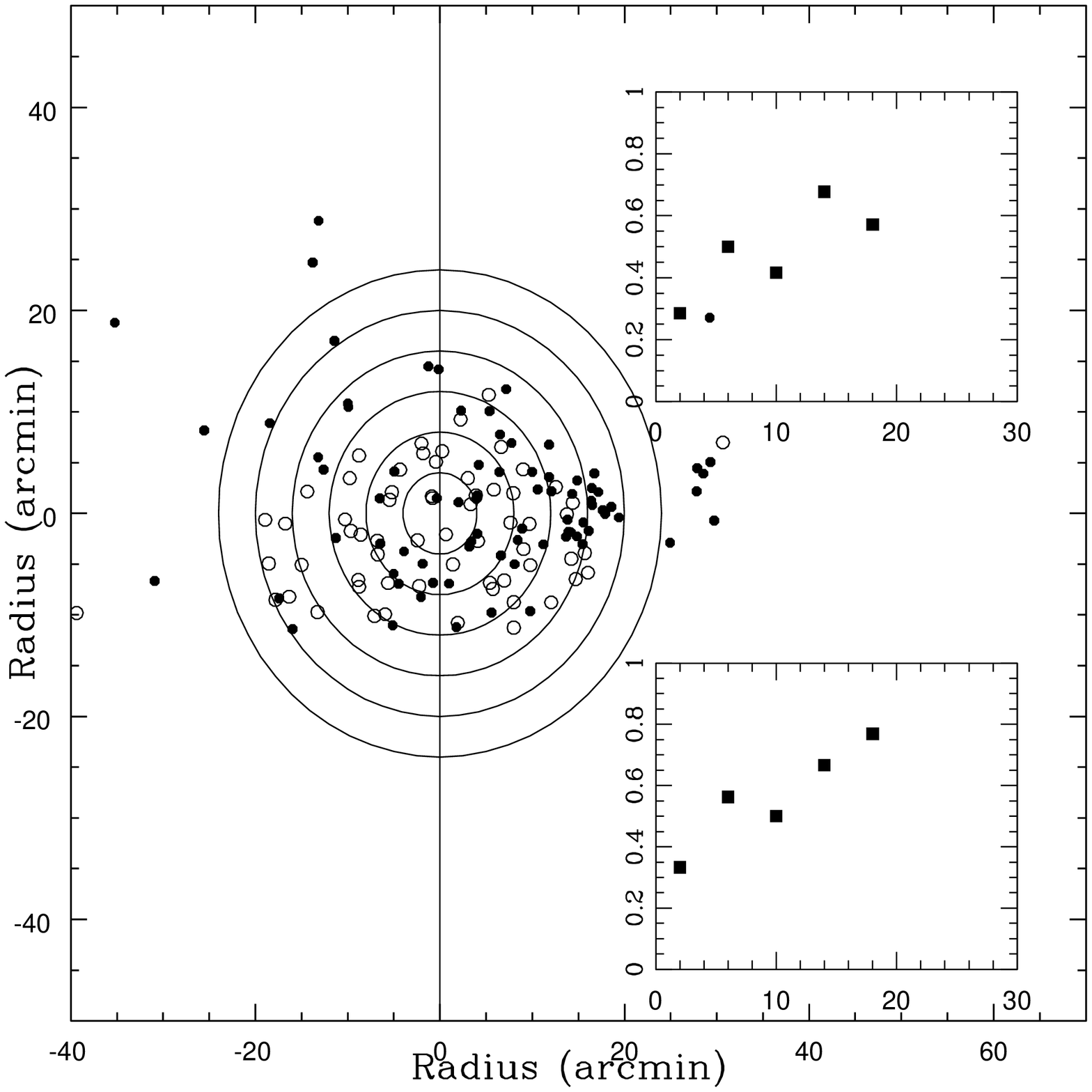}}
\caption{Left panel: The distribution of Class III (open circles) and Class II (filled circles) YSOs in NGC 1893 region. The concentric annuli around the center have a width of 1\arcm. The solid line demarcates the sample of YSOs towards and opposite direction of the nebuale. The upper and lower insets show variation of $f_{CTTS}$ as a function of radial distance from the cluster center towards the direction of the nebulae and for the whole cluster region respectively.   
Right panel:  Same as left panel but for Tr 37/ IC 1396 region.}
\label{fig9}
\end{figure*}

\subsection {Evolution of disk of TTSs}

As discussed in Sec 5, the YSOs selected in the present study have masses less than $\sim$ 3$\msun$, hence these YSOs are likely to  be TTSs. Recently, a considerable interest has been evolved to study the evolution of circumstellar disks around the YSOs (e.g., Haisch et al. 2001, Sicilia-Aguilar et al. 2006, Carpenter et al. 2006, Chauhan et al. 2009). The motivation to study the disk evolution lies in the fact that circumstellar disks are the potential sites of planet formation. In fact  the observed masses, sizes and chemical composition of young circumstellar disks are found to be analogous to that the protosolar nebula 
(e.g., Beckwith 1999, Hillenbrand 2005). Characterization of the disk evolution can improve the understanding of the planet formation. Different observational diagnostics, such as emissions lines, UV excess, NIR/ MIR excess, millimeter and sub-millimeter excess are being used to study the temporal evolution of circumstellar disks. These diagnostics trace different properties of the disks at different scales, e.g., emission lines and UV excess trace level of accretion activity, IR excess traces the hot inner disk, whereas millimeter and sub-millimeter excess trace cold dust in the outer disk. Most of the disk evolution studies available in the literature are based on the NIR/ MIR excess which traces the evolution of inner accretion disks and it is found that in majority of the cases the inner disk does not last for more than $\sim$ 6 Myr (see e.g., Haish et al. 2001). However, there are a few open questions such as whether disk material dissipates at all radii simultaneously or inner disk disappears first and propagates outwards, whether the dissipation of disk depends on stellar masses and whether the CTTSs evolve to WTTSs. 
Carpenter et al. (2006) have found that the evidence for mass  dependent circumstellar disk evolution, in the sense that the mechanism for disk dispersal operates less efficiently for low mass stars.

The ``standard model" by Kenyon \& Hartmann (1995) suggests that the CTTSs evolve to the WTTSs by losing the inner part of circumstellar disk. The age distribution of TTSs in the Taurus region indicates that the WTTSs are systematically older than the CTTSs, but the statistical significance was low (Kenyon \& Hartmann 1995; Hartmann 2001; Armitage et al. 2003). In the case of Taurus-Auriga T-association Bertout et al. (2007) concluded that the observed age distribution of CTTSs and WTTSs in the region can be explained by assuming that a CTTS evolves into a WTTS. Chauhan et al. (2009) have studied evolution of disks of TTSs associated with BRCs and supported the conclusion of Bertout et al. (2007) that CTTS evolves into a WTTS. On the other hand, there have also been many observations which claimed that the CTTS and the WTTS are coeval and have indistinguishable stellar properties (e.g., Walter et al. 1988; Lawson et al. 1996; Gras-Velazquez \& Ray 2005). From the analysis of the HR diagram of the CTTSs and WTTSs in Chamaeleon I, Lawson et al. (1996) concluded that some stars may be born even almost diskless or lose the disk at very early stages (age $<$ 1 Myr).  In such cases the coeval distribution of Class II sources (CTTSs) and Class III sources (WTTSs) does not invalidate the paradigm that CTTSs evolve to WTTSs.


\begin{figure}
\centering
\resizebox{12cm}{12cm}{\includegraphics{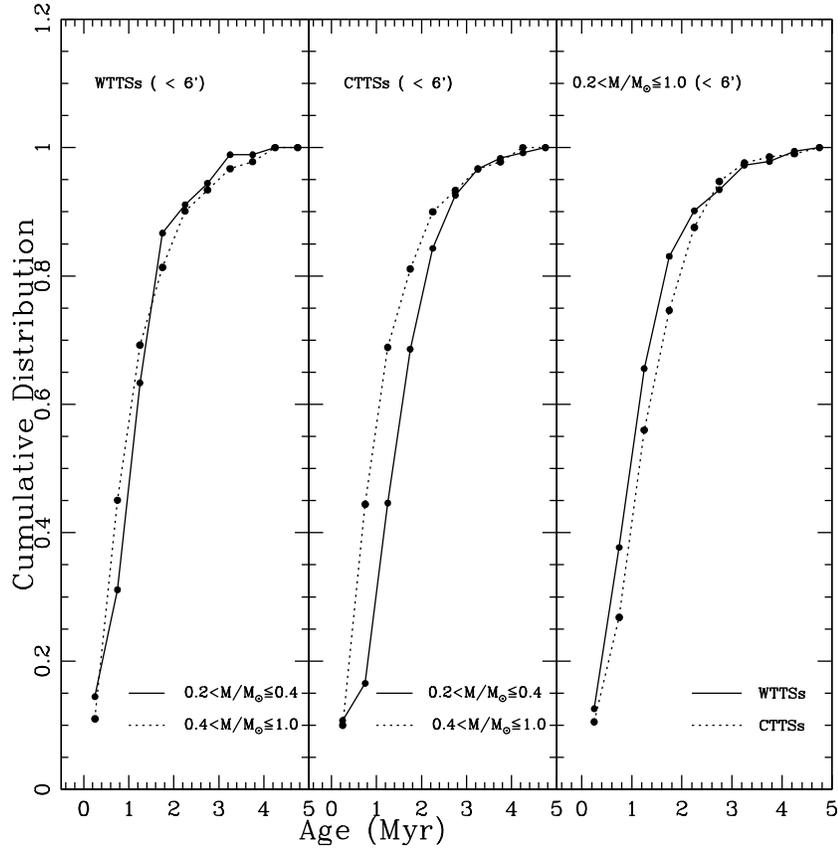}}
\caption{Left panel: Cumulative distributions of WTTSs of NGC 1893 region in two mass bins as a function of stellar age. Middle panel: Same as left panel but for CTTSs. Right Panel:  Cumulative distributions of CTTSs and WTTSs  in the mass range  0.2 $<$\msun $\le$ 1.0  as a function of stellar age}
\label{fig10}
\end{figure}

In the present work, we have estimated ages of 538 TTSs (see Sec. 5), hence the sample can be used to explore the answers to some of the above questions. In Fig. 10 we compare cumulative distribution of Class III (presuming as WTTSs; left panel) and Class II (presuming as CTTSs; middle panel) sources lying within the NGC 1893 cluster region and  having masses in the range 0.2 $<$ M/\msun $\le$ 0.4 and 0.4 $<$ M/\msun $\le$ 1.0, respectively. Fig. 10 (right panel) compares the cumulative distribution of Class II and and Class III sources having masses in the range of 0.2 $<$ M/\msun $\le$ 1.0. The cumulative age distributions of WTTSs and CTTSs in two mass bins (Fig. 10, left and middel panels respectively) indicates that the sources having masses $\sim 0.4 < M/M_\odot \le 1.0$ are relatively younger than WTTSs/ CTTSs having lower masses.  The Kolmogorov - Smirnov (KS) test confirms that the cumulative distributions of WTTSs and CTTSs in these two mass bins are different at a confidence levels of $\sim$ 80\% and $>99\%$ respectively.  This is a quantitative evidence supporting  the notion that the mechanisms for disk dispersal operates less efficiently for low-mass CTTSs. A similar result has been  derived on the  basis of disk fraction of TTSs in different mass bins by Carpenter et al. (2006) and references therein. 
Roccatagliata et al. (2011) have found that the disk fraction in IC 1795 also depends on the stellar mass. They found that the  sources with masses $>2 \msun$ have a disk fraction of $\sim 20\%$, while lower mass objects ($\sim 0.8 < M/M_\odot \le 2.0$) have a disk fraction of $\sim 50\%$ which implies that disks around massive stars have a shorter dissipation timescale.

The cumulative distributions of WTTSs and CTTSs in the mass range 0.2 $<$ M/\msun $\le$ 1.0 reveal that these two distributions are different at a confidence level of $\sim 90\%$ suggesting that the WTTSs are younger in comparison to the CTTSs. This result is in contradiction with that recently found by Chauhan et al. (2009) in the case of CTTSs and WTTSs (classified on the basis of H$\alpha$ EW) associated with six BRCs and by Bertout et al. (2007) in the case of Taurus-Auriga T association. However, the above result is in agreement with those which claim that the CTTSs and WTTSs are coeval and have indistinguishable properties (e.g., Walter et al. 1988, Lawson et al. 1996, Gras-Velazquez \& Ray 2005, Chauhan et al. 2011).  P11 also has reported that the age and mass distribuions of Class II and Class III sources in the NGC 1893 are similar.

\subsubsection {Comparison with IC 1396}

\begin{figure}
\centering
\resizebox{12cm}{12cm}{\includegraphics{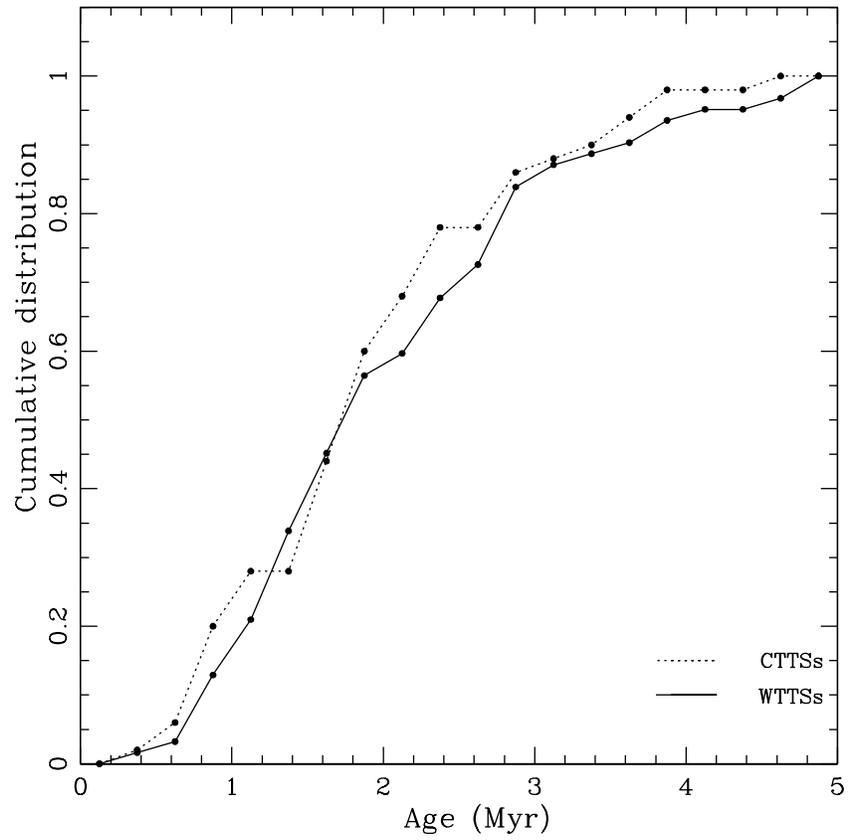}}
\caption{Cumulative distribution of CTTSs and WTTSs of Tr 37/ IC 1396 region in the mass range  0.3 $<$ M/\msun $\le$ 3.0 as a function of stellar age.
}
\label{fig11}
\end{figure}

Since the morphology and age of Tr 37/ IC 1396 is rather similar to that of NGC 1893, it will be worthwhile to compare the age distribution of CTTSs and WTTSs of NGC 1893 with those in the Tr 37/ IC 1396 globule region. The CTTSs and WTTSs in the Tr 37/ IC 1396 globule region are identified on the basis of EW of H$\alpha$ emission (stars having H$\alpha$ EW $\geq$ 10 {\AA}   are considered CTTSs). The data for EW and ages have been taken from Sicilia-Aguilar et al. (2005). Fig. 11 shows cumulative distributions of CTTSs and WTTSs of Tr 37 region having masses in the range $0.3 < M/M_\odot \le 3.0$  which indicates that although the distribution of WTTSs suggest higher ages for WTTSs, however the probability of these two samples belong to two different populations is only $\sim 30\%$.

\section{Summary}
On the basis of a comprehensive multi-wavelength study of the star forming region NGC 1893, we continued our attempt to understand the star formation scenario in and around the young cluster NGC 1893. The main results of the present work are enumerated as below:
\begin{enumerate}
\item The YSOs associated with the cluster region show an age spread of $\sim$ 5 Myr. The ages of the YSOs located near the nebulae at the periphery of the cluster have relatively smaller ages as compared to those located in the cluster region. The result is in accordance with our earlier result (S07, Maheswar et al. 2007) which support a triggered star formation towards the direction of the nebulae.

\item The density distribution of identified YSOs indicates an aligned distribution from the ionization source to the direction of the nebulae located at the periphery of the cluster. Since the cluster is not dynamically relaxed the alligned distribution of the YSOs may be an imprint of star formation process.

\item The fraction of disk bearing stars increases towards the periphery of the cluster. 

\item There is an evidence supporting the notion that the mechanisms for disk dispersal operate less efficiently for low-mass stars. 

\item The sample of Class II sources is found to be relatively older in comparison to that of Class III sources. This result is in contradiction with that recently found by Chauhan et al. (2009) in the case of CTTSs and WTTSs (classified on the basis of H$\alpha$ EW) associated with six BRCs and by Bertout et al. (2007) in the case of Taurus-Auriga T association. However, the above result is in agreement with those which claim that the CTTSs and WTTSs are coeval and have indistinguishable properties (e.g., Walter et al. 1988, Lawson et al. 1996, Gras-Velazquez \& Ray 2005).

\end{enumerate}

\section*{Acknowledgments}
The authors are thankful to the anonymous referee for useful comments. Part of the work was carried out by AKP during his visit to National Central University (Taiwan) under India-Taiwan collaborative program. AKP is thankful to the GITA, DST (India) and NSC (Taiwan) for the financial help. This publication makes use of the data products from the 2MASS, which is a joint project of the University of Massachusetts and the Infrared Processing and Analysis Center/California Institute of Technology.

\begin{table}
\caption{ Age and mass of YSOs (Class II and Class III) used in the present study. The complete table is available in electronic form.}
\begin{tabular}{cccc}
\hline
{RA} & {DEC}  & {Age}& {Mass}  \\
{J2000}  & {J2000} & {Myr} & {\msun}\\
\hline
\multicolumn{4}{l}{Class III sources}\\
80.850998 & 33.431671 & 0.11    $\pm$   0.01 & 1.60     $\pm$   0.01  \\
80.699333 & 33.444271 & 1.46    $\pm$   0.40 & 2.86     $\pm$   0.13  \\
80.760300 & 33.443619 & 1.48    $\pm$   0.31 & 2.67     $\pm$   0.06  \\
80.721466 & 33.521236 & 0.28    $\pm$   0.03 & 1.53     $\pm$   0.09  \\
80.668961 & 33.389114 & 3.23    $\pm$   0.68 & 2.27     $\pm$   0.11  \\
...       & ...       & ...                  & ...                   \\
...       & ...       & ...                  & ...                   \\
\multicolumn{4}{l}{ Class II sources}\\
80.725868 & 33.514194 & 0.12    $\pm$   0.02 & 2.14     $\pm$   0.06 \\
80.589729 & 33.480995 & 0.10    $\pm$   0.01 & 1.45     $\pm$   0.01 \\
80.685089 & 33.522739 & 0.10    $\pm$   0.00 & 1.35     $\pm$   0.01 \\
80.723045 & 33.523594 & 0.49    $\pm$   0.06 & 1.62     $\pm$   0.09 \\
...       & ...       & ...                  & ...                   \\
...       & ...       & ...                  &...                    \\
\hline
\end{tabular}
\end{table}

\begin{thebibliography}{99}
\bibitem{b01} Adams F. C., Hollenbach D., Laughlin G., Gorti U., 2004, ApJ, 611, 360
\bibitem{b02} Armitage P. J., Clarke C. J., Palla F., 2003, MNRAS, 342, 1139 
\bibitem{b03} Balog Z., Muzerolle J., Rieke G. H., Su K. Y. L., Young E. T., Megeath S. T., 2007, ApJ, 660, 1532
\bibitem{b04} Barentsen G., Vink J. S., Drew J. E., Greimel R., Wright N. J., Drake J. J., Martin E. L., Valdivielso L., Corradi R. L. M., 2011, MNRAS, 415, 103
\bibitem{b05} Beckwith S. V. W., 1999, in Lada C. J., KylafisN. D., eds, NATO ASIC Proc. 540, The Origin of Stars and Planetary Systems. Kluwer, Dordrecht  p. 579
\bibitem{b06} Bertoldi F., 1989, ApJ, 346, 735
\bibitem{b07} Bertout C., Siess L., Cabrit S. 2007, A\&A, 473, L21
\bibitem{b08} Bessell M.S., Brett J.M., 1988, PASP, 100, 1134
\bibitem{b09} Bouwman J., Lawson W. A., Dominik C., Feigelson E. D., Henning Th., Tielens A. G. G. M., Waters L. B. F. M., 2006, ApJ, 653, 57
\bibitem{b10} Caramazza M., Micela G., Prisinzano L., Rebull L., Sciortino S., Stauffer J. R., 2008, A\&A, 488, 211
\bibitem{b11} Carpenter J. M., Mamajek E. E., Hillenbrand L. A., Meyer M. R., 2006, ApJ, 651, L49
\bibitem{b12} Chauhan N., Pandey A. K., Ogura K., Ojha D. K., Bhatt B. C., Ghosh S. K., Rawat P. S., 2009, MNRAS, 396, 964
\bibitem{b13} Chauhan N., Pandey A. K., Ogura K., Jose, J., Ojha D. K., Samal, M. R., Mito, H., 2011, MNRAS, 415, 1202
\bibitem{b14} Chen W. P., Chen C. W., Shu C. G., 2004, AJ, 128, 2306
\bibitem{b15} Clarke C. J., 2007, MNRAS, 376, 1350
\bibitem{b16} Cohen J.G., Frogel J.A., Persson S.E., Ellias J.H., 1981, ApJ, 249, 481
\bibitem{b17} Cutri R. M. et al., 2003, 2MASS All Sky Catalog of Point Sources. NASA/IPAC Infrared Science Archive
\bibitem{b18} Dahm S. E., Simon T., 2005, AJ, 129, 829
\bibitem{b19} Deharveng L., Lefloch B., Zavagno A., Caplan J., Whitworth A. P., Nadeau D., Martin S., 2003, A\&A, 408, L25
\bibitem{b20} De Vries C.H., Narayanan G., Snell R.L., 2002, ApJ, 577, 798
\bibitem{b21} Elmegreen B. G., Lada C. J., 1977, ApJ, 214, 725
\bibitem{b22} Eswaraiah C., Pandey A. K., Maheswar G., Medhi B. J., Pandey J. C., Ojha D. K., Chen W. P., 2011, MNRAS, 411, 1418
\bibitem{b23} Girardi L., Bertelli G., Bressan A., Chiosi C., Groenewegen M.A.T., et al. 2002, A\&A, 391,195
\bibitem{b24} Gras-Velazquez A.,  Ray T. P., 2005, A\&A, 443, 541
\bibitem{b25} Haisch K.E., Lada E.A.,  Lada C.J., 2001, AJ, 121, 2065
\bibitem{b26} Hartmann L., 2001, AJ, 121, 1030
\bibitem{b27} Hillenbrand L. A., Strom S. E., Vrba F. J., Keene J., 1992, ApJ, 397, 613
\bibitem{b28} Hillenbrand L. A., Massey P., Strom S. E., Merrill K. M., 1993, AJ, 106, 1906
\bibitem{b29} Hillenbrand L. A., 2005, A Decade of Discovery: Planets Around Other Stars" STScI Symposium Series 19, ed. M. Livio, astro-ph/0511083
\bibitem{b30} Hillenbrand L. A., Bauermeister A., White R.J., 2008, in ASP Conf. Ser. 384, 14th Cambridge Workshop on Cool Stars, Stellar Systems, and the Sun, ed. G. van Belle (San Francisco, CA: ASP), 200
\bibitem{b31} Johnson H. L., Morgan W. W., 1953, ApJ, 117, 313
\bibitem{b32} Johnson H. L., 1966, ARA\&A, 4, 193
\bibitem{b33} Johnstone D., Hollenbach D., Bally J., 1998, ApJ, 499, 758
\bibitem{b34} Johnstone D., Matsuyama I., McCarthy I. G., Font A. S., 2004, R. Mex. A\&A, 22, 38
\bibitem{b35} Kenyon S.,  Hartmann L., 1995, ApJS, 101,117
\bibitem{b36} Kessel-Deynet O.,  Burkert A., 2003, MNRAS, 338, 545
\bibitem{b37} Koenig X.P., Allen L.E., Gutermuth R A., Hora J.L., Brunt C.M., Muzerolle J. 2008, ApJ, 688, 1142
\bibitem{b38} Lawson W. A., Fiegelson E. D., Huenemoerder D. P., 1996, MNRAS, 280, 1071
\bibitem{b39} Lefloch B., Lazareff B., 1995, A\&A, 301, 522
\bibitem{b40} Lefloch B., Lazareff B., Castets A. 1997, A\&A, 324, 249
\bibitem{b41} Maheswar G., Sharma S., Biman J. M., Pandey A. K., Bhatt H. C., 2007, MNRAS, 379, 1237
\bibitem{b42} Matsuyanagi I., Itoh Y., Sugitani K., Oasa Y., Mukai T., Tamura M., 2006, PASJ, 58, L29
\bibitem{b43} Meyer M., Calvet N., Hillenbrand, L. A., 1997, AJ, 114, 288
\bibitem{b44} Miao J., White G.J., Nelson R., Thompson M., Morgan L., 2006, MNRAS, 369, 143
\bibitem{b45} Ojha D. K. et al., 2004a, ApJ, 608, 797
\bibitem{b46} Ojha D. K. et al., 2004b, ApJ, 616, 1041
\bibitem{b47} Pandey A.K., Sharma S., Ogura K., Ojha D.K., Chen W.P., Bhatt B.C., Ghosh S.K. 2008, MNRAS, 383, 1241
\bibitem{b48} Prisinzano.  L., Sanz-Forcada.  J., G. Micela. G.,  Caramazza. M., Guarcello, M. G., Sciortino, S., \&  Testi, L., 2011, A\&A, 527, 77, (P11)
\bibitem{b49} Robitaille T.P., Whitney B.A., Indebetouw R., Wood K., Denzmore P., 2006, ApJS, 167, 256
\bibitem{b50} Roccatagliata V., Bouwman J., Henning T., Gennaro  M., Feigelson E., Kim J. S.,  Sicilia-Aguilar A., Lawson W. A., 2011, ApJ, 733, 113
\bibitem{b51} Samal M.R., Pandey A.K., Ojha D.K., Ghosh S.K., Kulkarni V.K., Bhatt B.C., 2007, ApJ, 671, 555
\bibitem{b52} Sharma S., Pandey A. K., Ogura K., Mito H., Tarusawa K., Sagar R., 2006, AJ, 132, 1669
\bibitem{b53} Sharma S., Pandey A.K., Ojha D.K., Chen W.P., Ghosh S.K., Bhatt B.C., Maheswar G., Sagar R., 2007, MNRAS, 380, 1141 (S07)
\bibitem{b54} Sharma, S., et al. 2012, PASJ, in press, arXiv 1204.2897
\bibitem{b55} Sicilia-Aguilar A., Hartmann L. W., Briceno C., Muzerolle J., Calvet N., 2004, AJ, 128, 805
\bibitem{b56} Sicilia-Aguilar A., Hartmann L. W., Harnandez J., Briceno C., Calvet N., 2005, AJ, 130, 188
\bibitem{b57} Sicilia-Aguilar A., Hartmann L. W., Fürész G., Henning T., Dullemond C., Brandner W., 2006, AJ, 132, 2135
\bibitem{b58} Siess L., Dufour E., Forestini M., 2000, A\&A, 358, 593
\bibitem{b59} Walter F. M., Brown A., Matthieu R. D., Meyer P. C., Vrba F. J., 1988, AJ, 96, 297 
\end{thebibliography}
\end{document}